\documentclass[10pt,a4j]{article}
\usepackage[dvips]{graphicx}
\usepackage{mathrsfs}
\usepackage{eucal}
\usepackage{amsmath,amsthm,amssymb}
\usepackage{wrapfig}
\usepackage[dvips]{color}
\usepackage{cite}
\usepackage{framed}
\DeclareMathVersion{chem}
\SetSymbolFont{letters}{chem}{OT1}{cmr}{m}{n}

\begin{document}

\renewcommand{\thefootnote}{$\dagger$\arabic{footnote}}

\begin{flushright}
\textit{Gel Formation in Drop-wise Addition Polymerization}
\end{flushright}
\vspace{1mm}

\begin{center}
\setlength{\baselineskip}{25pt}{\LARGE\textbf{Theory of Gel Formation \\{\Large Drop-wise Addition of R$-$A$_{f}$ solution onto R$-$B$_{g}$ Solution}}}
\end{center}
\vspace{0mm}

\vspace*{0mm}
\begin{center}
\large{Kazumi Suematsu} \vspace*{2mm}\\
\normalsize{\setlength{\baselineskip}{12pt} 
Institute of Mathematical Science\\
Ohkadai 2-31-9, Yokkaichi, Mie 512-1216, JAPAN\\
E-Mail: suematsu@m3.cty-net.ne.jp,  Tel/Fax: +81 (0) 593 26 8052}\\[8mm]
\end{center}

%%%%%%%%%%%%%%%%%%
\hrule
\vspace{0mm}
\begin{flushleft}
\textbf{\large Abstract}
\end{flushleft}
A theory of drop-wise addition polymerization is developed. Because of the linear growth of the reaction volume $V$, the system gives rise to a new type of distribution function for cyclic species that can be expressed by the sum of two terms: the conventional distribution term and a extra term due to the dilution effect. The present result is an extension of the conventional homogeneous polymerization that corresponds to a special case without the extra term. Making use of the result we derive the gel point formula for this unique polymerization. The theoretical result is compared with the recent observations.\\[-3mm]
\begin{flushleft}
\textbf{\textbf{Key Words}}:
\normalsize{Drop-wise Addition/ Distribution of Cyclic Species/ Gel Point}\\[3mm]
\end{flushleft}
\hrule
\vspace{3mm}

\setlength{\baselineskip}{13pt} 
\section{Introduction}
Organic chemists often make use of the technique of the drop-wise addition. When the synthesis of macrocyclic compounds is designed, the drop-wise addition has been employed frequently: '\textit{Together with use of a large amount of solvents a technique is introduced that a reagent is added slowly keeping the system in high dilution, since the molecule is more likely to react with itself than with other molecules in high dilution}.' In contrast, the same technique has rarely been employed in polymer chemistry where constructing intermolecular bonds in longer sequence is necessarily required; intramolecular bonds simply waste valuable functional units, thus giving useless, unwelcome cyclic by-products. 

Quite recently an interesting synthetic architecture of the drop-wise addition polymerization was put forth by Turkish group\cite{Yilgor}. It has been known earlier that aliphatic isocyanates (-NCO) react with aliphatic amines (-NH$_{2}$) more than one thousand times faster than with alcohols (-OH). Making full use of this velocity difference between amines and alcohols, Yilgor and coworkers investigated the polymerizability of diisocyanates and diamines in isopropyl alcohol (IPA). Typically an IPA solution of a diisocyanate and that of a diamine are prepared separately; the diamine solution is introduced into the reaction vessel and the diisocianate solution the addition funnel; then the diisocianate solution is slowly added drop-wise onto the diamine solution. They found that the reaction is very fast at room temperature and yields clear and homogeneous polymer solutions of high molecular weight. 

To date no theoretical treatment has been put forth for the drop-wise addition polymerization. The purpose of this paper is to construct the theory for this interesting polymerization. First we discuss the molecular size distribution of linear polymers without rings using kinetic arguments, then extend the result to the branched polymers. It has been shown that the size distribution for tree clusters can be derived beautifully by the statistical-mechanics approach\cite{Good}. In order to to gain deep insights into the drop-wise addition polymerization, however, the kinetic treatment is essential and is found to be powerful. We show that the resultant distribution function is equivalent to that of the multiple link system\cite{Tanaka}. Next we derive the size distribution of cyclic species, showing that because of the continuous change of the system volume in the drop-wise addition polymerization, a new type of distribution function arises. Making use of the result together with the known formalism\cite{Kazumi}, finally we derive the gel point equation. The theoretical result is compared with recent observed data.

%%%%%%%%%%%%%%%%%% Section 1
\section{Theoretical}
The basic assumption of the drop-wise addition polymerization is that
\begin{enumerate}
\item a solution of R$-$A$_{f}$ molecules is added drop-wise into a solution of R$-$B$_{g}$ molecules;
\item the velocity of reaction is sufficiently slow so that the complete mixing is achieved before the reaction starts; as a result the principle of equireactivity is fulfilled;
\begin{flushleft}
\text{quite conversely}
\end{flushleft}
\item the velocity of reaction must be sufficiently fast in the time scale of the dropping interval ($\delta u$), so that all the R$-$A$_{f}$ molecules contained in one droplet must be consumed completely forming A$-$B bonds within $\delta u$. As a result the system is always comprised of the $\left(\text{BB}-\text{AA}\right)_{n}\text{BB}$ type molecular species alone \hspace{1mm}($n$ represents the number of R$-$A$_{f}$ molecules and $n=0,1,2,\cdots$). 
\end{enumerate}

\subsection{Ideal Drop-wise Addition Polymerization}
We consider the limiting case that each droplet contains only one molecule; namely,\textit{a single} R$-$A$_{f}$  molecule is added drop-wise onto the solution of the R$-$B$_{g}$ monomers. 

Consider the transition from $u$ to $u+\delta u$ drops $(\delta u=1)$. In this minute interval, the following set of elementary reactions should occur.
%%%%%%%%%%%%%%%%%% Eq. (1)
\begin{align}
\left(\text{BB}-\text{AA}\right)_{x-1}\text{BB}\hspace{3mm} +\hspace{3mm} \text{AA}\hspace{5mm} &\rightarrow\hspace{5mm} \left(\text{BB}-\text{AA}\right)_{x}\\
%%%%%%%%%%%%%%%%%% Eq. (2)
\left(\text{BB}-\text{AA}\right)_{j}\hspace{3mm} +\hspace{3mm} \left(\text{BB}-\text{AA}\right)_{x-j-1}\text{BB}\hspace{5mm} &\rightarrow\hspace{5mm} \left(\text{BB}-\text{AA}\right)_{x-1}\text{BB}\\
%%%%%%%%%%%%%%%%%% Eq. (3)
\left(\text{BB}-\text{AA}\right)_{x}\hspace{5mm} &\rightarrow\hspace{5mm}\text{ring}-\left(\text{BB}-\text{AA}\right)_{x}
\end{align}
with $x=1,2,3, \cdots$. Eq. (2) (intermolecular reaction) competes with eq. (3) (cyclization reaction). An essential point is that the above set of reactions (1)-(3) must be completed within $\delta u$. So these reactions constitute a unit chemical cycle. The drop-wise addition polymerization proceeds repeating this cycle.
\begin{flushleft}
\subsubsection{Molecular Size Distribution in Linear Process}
\end{flushleft}
First consider the linear system without rings. Let $M_{B}$ denote the total number of the R$-$B$_{g}$ units in the system. Let BA$_{x}$ and BB$_{x-1}$ be the abbreviations of the molecular species, (BB$-$AA)$_{x}$ and (BB$-$AA)$_{x-1}$BB, respectively. Then their variations are
for eq. (1)
%%%%%%%%%%%%%%%%%% Eq. (4)
\begin{align}
\delta N_{BA_{x}}&=\frac{2N_{BB_{x-1}}\cdot 2N_{AA}}{2M_{B}(1-D_{B})\cdot 2N_{AA}}\hspace{5mm} \text{for} \hspace{1mm} \left(\text{BB}-\text{AA}\right)_{x}, \\
%%%%%%%%%%%%%%%%%% Eq. (5)
\delta N_{BB_{x-1}}&=\frac{-2N_{BB_{x-1}}\cdot 2N_{AA}}{\displaystyle2M_{B}(1-D_{B})\cdot 2N_{AA}}\hspace{5mm}\text{for} \hspace{1mm}  \left(\text{BB}-\text{AA}\right)_{x-1}\text{BB},
\end{align}
and for eq. (2)
%%%%%%%%%%%%%%%%%% Eq. (6)
\begin{equation}
\delta N_{BB_{x-1}}=\frac{\displaystyle\sum\nolimits_{j=1}^{x-1}\delta N_{BA_{j}}\cdot 2N_{BB_{x-j-1}}-2N_{BB_{x-1}}\sum\nolimits_{k=1}^{\infty}\delta N_{BA_{k}}}{\displaystyle2M_{B}(1-D_{B})\sum\nolimits_{k=1}^{\infty}\delta N_{BA_{k}}}.
\end{equation} 
By the definition of the ideal drop-wise addition, $N_{AA}=1$ and $\sum_{k=1}^{\infty}\delta N_{BA_{k}}=1$. Combining eqs. (4) and (5) with eq. (6), we have
%%%%%%%%%%%%%%%%%% Eq. (7)
\begin{equation}
\delta N_{BB_{x-1}}=\frac{\displaystyle\sum\nolimits_{j=1}^{x-1}\textstyle\left\{\frac{N_{BB_{j-1}}}{M_{B}(1-D_{B})}\right\}\cdot 2N_{BB_{x-j-1}}-4N_{BB_{x-1}}}{2M_{B}(1-D_{B})}\hspace{1mm}\delta u.
\end{equation}
Eq. (7) may be recast in the familiar form:
%%%%%%%%%%%%%%%%%% Eq. (8)
\begin{equation}
\delta N_{BB_{x-1}}=\frac{\frac{1}{2}\displaystyle\sum\nolimits_{j=1}^{x-1}\hspace{0.3mm}N_{BB_{j-1}}\hspace{0.3mm}\hspace{0.3mm}N_{BB_{x-j-1}}-N_{BB_{x-1}}\hspace{0.3mm}M_{B}(1-D_{B})}{\frac{1}{2}\Big\{M_{B}(1-D_{B})\Big\}^{2}}\hspace{1mm}\delta u.
\end{equation}
The (BB$-$AA)$_{x-1}$BB unit is produced by way of the formation of two bonds. So $D_{B}=2u/2M_{B}$. Now eq. (8) is soluble by means of the sequential operation:\\
1. \textit{for} $x=1$
$$\delta N_{BB_{0}}=\frac{-2N_{BB_{0}}}{(1-D_{B})}\hspace{0.6mm}\delta D_{B},$$
which yields
%%%%%%%%%%%%%%%%%% Eq. (9)
\begin{equation}
N_{BB_{0}}=M_{B}\left(1-D_{B}\right)^{2}.
\end{equation}
2. \textit{for}  $x=j$\\
Let the equation
%%%%%%%%%%%%%%%%%% Eq. (10)
\begin{equation}
N_{BB_{j-1}}=M_{B}D_{\hspace{-0.3mm}B}^{\hspace{0.8mm}j-1}(1-D_{B})^{2}
\end{equation}
be true for $x=j\left(\ge 2\right)$. Then substituting eq. (10) into eq. (8), we have
%%%%%%%%%%%%%%%%%% Eq. (11)
\begin{equation}
\delta N_{BB_{x-1}}+\frac{2N_{BB_{x-1}}}{1-D_{B}}\hspace{0.6mm}\delta D_{B}=(x-1)M_{B}D_{B}^{x-2}\left(1-D_{B}\right)^{2}\hspace{0.6mm}\delta D_{B}.
\end{equation}
Multiply both sides of eq. (11) by the integrating factor $\lambda=\left(1-D_{B}\right)^{-2}$ to yield
%%%%%%%%%%%%%%%%%% Eq. (12)
\begin{equation}
\left(1-D_{B}\right)^{-2}\delta N_{BB_{x-1}}+\frac{2N_{BB_{x-1}}}{\left(1-D_{B}\right)^{3}}\hspace{0.6mm}\delta D_{B}=(x-1)M_{B}D_{B}^{x-2}\hspace{0.6mm}\delta D_{B}.
\end{equation}
Now our equation is exact. Integrating eq. (12), we have
%%%%%%%%%%%%%%%%%% Eq. (13)
\begin{equation}
N_{BB_{x-1}}=M_{B}D_{\hspace{-0.3mm}B}^{\hspace{0.8mm}x-1}(1-D_{B})^{2}.
\end{equation}
Thus eq. (13) is true for all $x$'s. The probability of finding $(x-1)$-mers is then
%%%%%%%%%%%%%%%%%% Eq. (14)
\begin{equation}
p_{BB_{x-1}}=\frac{N_{BB_{x-1}}}{\sum\nolimits_{x=1}^{\infty}N_{BB_{x-1}}}=D_{\hspace{-0.3mm}B}^{\hspace{0.8mm}x-1}(1-D_{B}),
\end{equation}
where $x=1,2,3, \cdots$.

It is more convenient to recast eq. (14) in the form:
%%%%%%%%%%%%%%%%%% Eq. (15)
\begin{equation}
p_{BB_{n}}=D_{\hspace{-0.3mm}B}^{\hspace{0.8mm}n}(1-D_{B}), \hspace{3mm}\left(n=0, 1,2,3, \cdots\right).
\end{equation}
Now eq. (15) represents the distribution of the (BB$-$AA)$_{n}$BB molecules having $n$ AA units.\\

\subsubsection{Molecular Size Distribution in Branching Process}
We seek the distribution function of the (BB$-$AA)$_{n}$BB type branched molecules having $n$ R$-$A$_{f}$ monomers ($n=0,1,2,\cdots$). Consider a tree molecule without rings. This molecule comprises
%%%%%%%%%%%%%%%%%% Eq. (16)
\begin{align}
&\text{number of A type molecules}:& &n\nonumber\\
&\text{number of B type molecules}:& &\{(f-1)n+1\}\nonumber\\
&\text{number of unreacted A FU's}:& &0\nonumber\\
&\text{number of unreacted B FU's}:& &\nu_{n}=g+\{(f-1)(g-1)-1\}n\nonumber\\
&\text{number of reacted A FU's}:& &fn\left(\equiv\text{number of reacted B FU's}\right).
\end{align}
General birth-death formula for \textit{n}-mers is

%%%%%%%%%%%%%%%%%% Eq. (17)
\begin{equation}
\delta N_{n}/\delta u=P_{birth}-P_{death}.
\end{equation}
The first term of the right hand side represents the birth probability of \textit{n}-mers and the second term the death probability.
The birth-death equation has the analytic expression of the form\,\footnote{\,To check the validity of eq. (18), carry out the summation over all molecular species from $n=0$ to $\infty$ to yield: $$\delta\Omega_{0}=\delta\sum\nolimits_{n=0}^{\infty}N_{n}=-(f-1)\delta u,$$
which satisfies the Euler relation: $\Omega_{0}=M_{B}-(f-1)u$, in support of eq. (18).}:

%%%%%%%%%%%%%%%%%% Eq. (18)
\begin{equation}
\frac{\delta N_{n}}{\delta u}=\frac{\frac{1}{f!}\displaystyle\sum\nolimits_{\{k_{\ell}\}=0}^{n-1}\prod\nolimits_{\ell=1}^{f}\nu_{k_{\ell}}N_{k_{\ell}}}{\frac{1}{f!}\left\{gM_{B}(1-D_{B})\right\}^{f}}\hspace{1mm}-\hspace{1mm}\frac{\frac{1}{(f-1)!}\,\nu_{n}N_{n}\cdot \left\{gM_{B}(1-D_{B})\right\}^{f-1}}{\frac{1}{f!}\left\{gM_{B}(1-D_{B})\right\}^{f}},
\end{equation}
where the summation of the first term is over all combinations that satisfies $k_{1}+k_{2}+\cdots +k_{f}=n-1$ along with $0\le k_{\ell}\le n-1\,\,(\ell=1, 2,\cdots,f)$. For instance, for the simplest case of $f=2$, eq.(18) reduces to the familiar form:

$$\frac{\delta N_{n}}{\delta u}=\frac{\frac{1}{2}\displaystyle\sum\nolimits_{j=0}^{n-1}\nu_{j}N_{j}\cdot\nu_{n-j-1}N_{n-j-1}}{\frac{1}{2}\left\{gM_{B}(1-D_{B})\right\}^{2}}\hspace{1mm}-\hspace{1mm}\frac{\nu_{n}N_{n}\cdot \left\{gM_{B}(1-D_{B})\right\}}{\frac{1}{2}\left\{gM_{B}(1-D_{B})\right\}^{2}}.$$

\noindent Using the equality, $fu/gM_{B}=D_{B}$, it is easy to show that $N_{0}=M_{B}(1-D_{B})^{\nu_{0}}$. We expect the solution of eq. (18) is generally of the form:
%%%%%%%%%%%%%%%%%% Eq. (19)
\begin{equation}
N_{k}=\frac{M_{B}}{(f-1)k+1}\hspace{0.5mm}\omega_{k}D_{B}^{\,k}(1-D_{B})^{\nu_{k}},
\end{equation}
where $$\omega_{k}=g\frac{\left(\nu_{k}+k-1\right)!}{k!\cdot\nu_{k}!}\,\,\,\,(k=0,1,2, \cdots).$$

\noindent Assume that eq. (19) is true for a given \textit{k}. Then substituting this equation into eq. (18) we have
%%%%%%%%%%%%%%%%%% Eq. (20)
\begin{equation}
\delta N_{n}+\frac{\nu_{n}N_{n}}{1-D_{B}}\delta D_{B}=\frac{g}{f}M_{B}\hspace{-1mm}\sum_{\{k_{\ell}\}=0}^{n-1}\prod_{\ell=1}^{f}\frac{1}{(f-1)k_{\ell}+1}\nu_{k_{\ell}}\omega_{k_{\ell}}D_{B}^{\,n-1}(1-D_{B})^{\nu_{n}}\,\delta D_{B}.
\end{equation}
Multiplying eq. (20) by the integrating factor, $\lambda=(1-D_{B})^{-\nu_{n}}$, and with the help of the equality
%%%%%%%%%%%%%%%%%% Eq. (21)
\begin{equation}
\frac{1}{f}\sum_{\{k_{\ell}\}=0}^{n-1}\prod_{\ell=1}^{f}\frac{1}{(f-1)k_{\ell}+1}\binom{\nu_{k_{\ell}}+k_{\ell}-1}{k_{\ell}}\equiv\frac{1}{(f-1)n+1}\binom{\nu_{n}+n-1}{n-1},
\end{equation}
we have
%%%%%%%%%%%%%%%%%% Eq. (19')
$$
N_{n}=\frac{M_{B}}{(f-1)n+1}\hspace{0.5mm}\omega_{n}D_{B}^{\,n}(1-D_{B})^{\nu_{n}},\eqno (19')
$$
which is just eq. (19). Thus, since eq. (19) was true for $k=0$, it is true for all $k$'s ($k=0,1,2,\cdots$). 
We see that eq. (19) is equal to  the cluster distribution function for the multiple link system\cite{Tanaka,Kazumi}, if we simply replace $f$ (functionality of the A type monomer) with $J$ (number of junction points). And for $f=2$ eq. (19) reduces to the known formula of the R$-$A$_{f}$ model. This is the reason why no one has so far addressed the theory of the drop-wise addition polymerization. The analogy between the drop-wise addition polymerization and the multiple link system is, however, only superficial. The situation changes drastically when one takes into consideration the formation of rings.

%%%%%%%%%%%%%%%%%% Fig.1
\begin{wrapfigure}[5]{r}{5.5cm}
\vspace*{-14mm}
\begin{center}
\includegraphics[width=5.3cm]{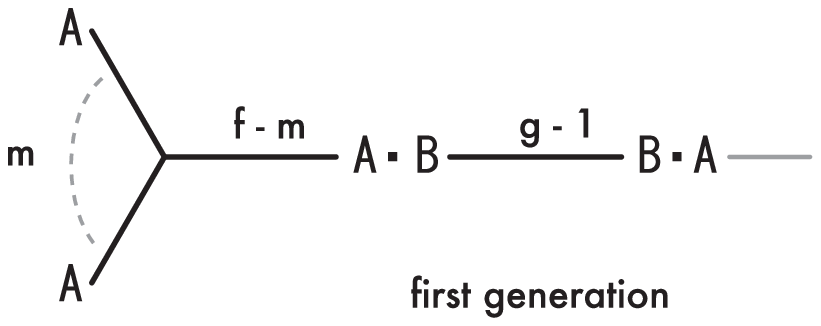}\\[3mm]
\end{center}
\setlength{\baselineskip}{10.5pt}{\small  Fig. 1: Representation of an m-tree of the R$-$A$_{f}\hspace{0.5mm}+\hspace{0.5mm}$R$-$B$_{g}$ branching model.}
\end{wrapfigure}

%%%%%%%%%%%%%%%%%% Section
\subsubsection{Distribution of Cyclic Species}
Let $C_{f,i}=fM_{A}/V_{A}$ be the initial functional unit (FU) concentration of the R$-$A$_{f}$ monomer solution before mixing and  $C_{g,i}=gM_{B}/V_{B}$ the corresponding quantity of the R$-$B$_{g}$ monomer solution. In concentrated solutions, we can approximate that all clusters are made up from the tree structure.
Suppose an m-tree which has \textit{m} unreacted A FU's on the root (see Fig. 1). The number of B FU's is then for the $x$th generation on this tree
%%%%%%%%%%%%%%%%%%
\begin{equation}
N(B)_{x}=(f-m)(g-1)\left[(f-1)(g-1)D_{B}\right]^{x-1}.\nonumber
\end{equation}
So, the number of chances, $\phi_{x,m}$, of $x$-ring formation is
%%%%%%%%%%%%%%%%%% Eq. (22)
\begin{equation}
\phi_{x,m}=m\times (f-m)(g-1)\left[(f-1)(g-1)D_{B}\right]^{x-1}(1-D_{B}).
\end{equation}
%%%%%%%%%%%%%%%%%%

Let $\mathscr{P}$ be the probability that one end on an \textit{x}-chain enters the small volume $v$ around another end on the same chain. Then the velocity of \textit{x}-ring formation is $\varpropto\mathscr{\hspace{-1mm}P}\phi_{x,m}$, while the velocity of intermolecular reaction is $\varpropto gM_{B}(1-D_{B})\cdot m\left(v/V\right)$. In concentrated solutions, the fraction of an $x$-ring to be formed for this special tree can be approximated by the relative velocity of the form\cite{Kazumi}:
%%%%%%%%%%%%%%%%%% Eq. (23)
\begin{equation}
\delta N_{R_{x,m}}\cong\frac{\mathscr{P}\hspace{0.6mm}m\hspace{0.3mm}(f-m)(g-1)\left[(f-1)(g-1)D_{B}\right]^{x-1}(1-D_{B})}{gM_{B}(1-D_{B})\cdot m\left(v/V\right)}\hspace{1mm}\delta u.
\end{equation}
The total fraction is
%%%%%%%%%%%%%%%%%% Eq. (24)
\begin{equation}
\sum_{x=1}^{\infty}N_{R_{x}}=\sum_{x=1}^{\infty}\,\int_{u}\,\sum_{m=1}^{f-1}\delta N_{R_{x,m}}=\sum_{x=1}^{\infty}\,\int_{0}^{u}V\hspace{0.3mm}\frac{f(f-1)}{2}\frac{\varphi_{x}(g-1)\left[(f-1)(g-1)D_{B}\right]^{x-1}(1-D_{B})}{gM_{B}(1-D_{B})}\hspace{1mm}du,
\end{equation}
where $\varphi_{x}=\mathscr{P}/v$. An essential point is that the system volume, $V$, varies with the advancement of reaction; i.e., the system is diluted successively with the addition of the R$-$A$_{f}$ solution. Since the volume element to be added during the unit interval ($\delta u=1$) is, by definition, $\delta V=V_{A}/M_{A}$ and $fu/gM_{B}=D_{B}$, the total volume at $u$ is 
%%%%%%%%%%%%%%%%%% Eq. (25)
\begin{equation}
V=\int_{0}^{u}\left({V_{A}}/{M_{A}}\right)du=\left(\frac{gM_{B}}{fM_{A}}D_{B}\right)V_{A}+V_{B},
\end{equation}
From eqs. (24) and (25) together with $fdu/gM_{B}=dD_{B}$, we have
%%%%%%%%%%%%%%%%%% Eq. (26)
\begin{equation}
\left[\varGamma\right]_{C\rightarrow\infty}=\sum_{x=1}^{\infty}N_{R_{x}}/V=\chi_{1}\sum_{x=1}^{\infty}\hspace{0.5mm}\varphi_{x}\frac{1}{2(x+1)}\left[(f-1)(g-1)D_{B}\right]^{x}+\chi_{2}\sum_{x=1}^{\infty}\hspace{0.5mm}\varphi_{x}\frac{1}{2x}\left[(f-1)(g-1)D_{B}\right]^{x},
\end{equation}
where 
%%%%%%%%%%%%%%%%%%
$$
\chi_{1}=\frac{\gamma_{f,i}D_{B}}{\gamma_{g,i}+\gamma_{f,i}D_{B}}\,\,\, \text{and}\,\,\,
\chi_{2}=\frac{\gamma_{g,i}}{\gamma_{g,i}+\gamma_{f,i}D_{B}}
$$
with $\gamma_{f,i}\, (=C_{f,i}^{\,-1})$ and $\gamma_{g,i}\, (=C_{g,i}^{\,-1})$ denoting the initial inverse-concentration of respective FU's before mixing. The first term of the right hand side in eq. (26) represents the extra term due to the dilution effect of the drop-wise addition. Hence the conventional homogeneous polymerization is a special case of $\chi_{1}=0$.

\subsection{General Drop-wise Addition Polymerization}
A more realistic model is that each droplet contains a mass of the R$-$A$_{f}$ monomer.  Let every droplet contain $L$ R$-$A$_{f}$ molecules ($L$ is a large number) which are injected drop-wise onto the R$-$B$_{g}$ solution. And consider the case $L\ll N_{A}$ (Avogadro number), so that each of the R$-$A$_{f}$ molecules reacts independently of the others. Assume that the mixing is complete so that the principle of equireactivity is assured. 

\subsubsection{Size Distribution of Cyclic Species}
Consider again a tree with \textit{m}-unreacted A FU's on the root. The number of B FU's in the $x$th generation after $k$ drops is then
%%%%%%%%%%%%%%%%%% Eq. (27)
\begin{equation}
N(B)_{x}=(f-m)(g-1)\left[(f-1)(g-1)D_{B}\right]^{x-1},
\end{equation}
where
%%%%%%%%%%%%%%%%%% Eq. (28)
\begin{equation}
D_{B}=\frac{(k-1)fL+f\varepsilon}{gM_{B}}\hspace{7mm} (0\le\varepsilon\le L, \,\text{and}\,\, k=1,2, \cdots).
\end{equation}
Thus $\varepsilon$ expresses the number of R$-$A$_{f}$ molecules that reacted in the interval from $k-1$ to $k$. The number of chances, $\phi_{x}$, of $x$-ring formation is thus for the \textit{m}-tree in question
%%%%%%%%%%%%%%%%%% Eq. (29)
\begin{equation}
\phi_{x,m}=m\times (f-m)(g-1)\left[(f-1)(g-1)D_{B}\right]^{x-1}(1-D_{B}).
\end{equation}
The rate of the ring formation can be expressed in the form:
%%%%%%%%%%%%%%%%%% Eq. (30)
\begin{equation}
v_{R}\propto \mathscr{P}\sum_{x=1}^{\infty}m\times (f-m)(g-1)\left[(f-1)(g-1)D_{B}\right]^{x-1}(1-D_{B})
\end{equation}
while the rate of intermolecular reaction is
%%%%%%%%%%%%%%%%%% Eq. (31)
\begin{equation}
v_{L}\propto m\times gM_{B}(1-D_{B})(v/V).
\end{equation}
In concentrated solutions, the fraction of rings to be formed per unit reaction $(\delta\varepsilon=1)$ may be approximated as $\delta N_{R}\cong v_{R}/v_{L}$\cite{Kazumi}. By eq. (28), we have $f\delta\varepsilon=gM_{B}\,\delta D_{B}$, which leads to
%%%%%%%%%%%%%%%%%% Eq. (32)
\begin{align}
\delta N_{R}&=(V/f)\sum_{m=1}^{f-1}\sum_{x=1}^{\infty}\varphi_{x}(f-m)(g-1)\left[(f-1)(g-1)D_{B}\right]^{x-1}\,\delta D_{B},\nonumber\\
&=(V/2)\sum_{x=1}^{\infty}\varphi_{x}(f-1)(g-1)\left[(f-1)(g-1)D_{B}\right]^{x-1}\,\delta D_{B}.
\end{align}
where $\varphi_{x}=\mathscr{P}/v$. Let $\lambda=fL/gM_{B}$. Then, with the equality $V=\frac{kL}{M_{A}}V_{A}+V_{B}$  in mind, integrate eq. (32)  from $D_{B}=(k-1)\lambda$ to $k\lambda$ to yield
%%%%%%%%%%%%%%%%%% Eq. (33)
\begin{align}
\Delta N_{R}(k)&=(V/2)\sum_{x=1}^{\infty}\int_{(k-1)\lambda}^{k\lambda}\varphi_{x}(f-1)(g-1)\left[(f-1)(g-1)D_{B}\right]^{x-1}\,dD_{B}\nonumber\\
&=\left(\frac{kL}{M_{A}}V_{A}+V_{B}\right)\left.\sum_{x=1}^{\infty}\varphi_{x}\frac{1}{2x}\left[(f-1)(g-1D_{B}\right]^{x}\right|_{(k-1)\lambda}^{k\lambda}\nonumber\\
&=\sum_{x=1}^{\infty}\varphi_{x}\frac{1}{2x}\left[(f-1)(g-1)\lambda\right]^{x}s(k),
\end{align}
where
%%%%%%%%%%%%%%%%%% Eq. (34)
\begin{equation}
s(k)=\left(\frac{kL}{M_{A}}V_{A}+V_{B}\right)\left\{k^{x}-(k-1)^{x}\right\}.
\end{equation}
The total number of rings accumulated from $k=1$ to $n$ drops is therefore
%%%%%%%%%%%%%%%%%% Eq. (35)
\begin{equation}
N_{R}(n)=\sum_{k=1}^{n}\Delta N_{R}(k)=\sum_{x=1}^{\infty}\varphi_{x}\frac{1}{2x}\left[(f-1)(g-1)\lambda\right]^{x}\mathscr{S}(n).
\end{equation}
Here, the inner sum of the r.h.s. has the form: 
%%%%%%%%%%%%%%%%%% Eq. (36)
\begin{align}
\mathscr{S}(n)&=\sum_{k=1}^{n} s(k)=\frac{L}{M_{A}}\left\{n^{x+1}-\left(0^{x}+1^{x}+2^{x}+\cdots+(n-1)^{x}\right)\right\}V_{A}\,+\,n^{x}\,V_{B}\nonumber\\
&=\frac{L}{M_{A}}n^{x}\left(n-\sum_{k=0}^{n-1}\left(k/n\right)^{x}\right)V_{A}\,+\,n^{x}\,V_{B}.
\end{align}
\\[3mm]

%%%%%%%%%%% Remark 1
\colorbox[gray]{0.90}{
\begin{minipage}{0.94\textwidth}
\paragraph{Remark 1:}
As $n\rightarrow \infty$, eq. (36) can be approximated as
%%%%%%%%%%% Eq. (37)
\begin{equation}
\mathscr{S}(n)\fallingdotseq\frac{L}{M_{A}}n^{x}\left(n-\int_{0}^{n}(k/n)^{x}dk\right)V_{A}\,+\,n^{x}\,V_{B}=\frac{nL}{M_{A}}\frac{x}{x+1}n^{x}V_{A}\,+\,n^{x}\,V_{B}.
\end{equation}
Substituting eq. (37) into  eq. (35), we have
%%%%%%%%%%% Eq. (38)
\begin{equation}
N_{R}(n)\fallingdotseq\frac{nL}{M_{A}}V_{A}\sum_{x=1}^{\infty}\varphi_{x}\frac{1}{2(x+1)}\left[(f-1)(g-1)n\lambda\right]^{x}+V_{B}\sum_{x=1}^{\infty}\varphi_{x}\frac{1}{2x}\left[(f-1)(g-1)n\lambda\right]^{x}.
\end{equation}
Using the equality, $n\lambda=D_{B}$, divide eq. (38) by $V(n)=\frac{nL}{M_{A}}V_{A}+V_{B}$ to yield
%%%%%%%%%%% Eq. (26')
$$
\left[\varGamma\right]\equiv\frac{N_{R}(n)}{V(n)}\fallingdotseq\chi_{1}\sum_{x=1}^{\infty}\varphi_{x}\frac{1}{2(x+1)}\left[(f-1)(g-1)n\lambda\right]^{x}+\chi_{2}\sum_{x=1}^{\infty}\varphi_{x}\frac{1}{2x}\left[(f-1)(g-1)n\lambda\right]^{x}, \eqno (26')
$$
which is just eq. (26). For a large $n$ limit, the general drop-wise addition polymerization converges on the ideal drop-wise addition polymerization.
\end{minipage}}\\[5mm]

%%%%%%%%%%% Remark 2
\colorbox[gray]{0.90}{
\begin{minipage}{0.94\textwidth}
\paragraph{Remark 2:}
The physical meaning of eq. (26) is as follows: 
\begin{enumerate}
\item{ If $\gamma_{f,i}\gg \gamma_{g,i}$}\\
The R$-$B$_{g}$ monomer is, for instance, in non-solvent state, into which the large amount of the R$-$A$_{f}$ dilute solution is injected. Thus cyclization occurs mainly due to the dilution effect by  the R$-$A$_{f}$ solution: $\chi_{1}=1$ and $\chi_{2}=0$.
 \item{If $\gamma_{g,i}\gg \gamma_{f,i}$}\\
The R$-$B$_{g}$ monomer is in dilution state, onto which a small volume of the concentrated R$-$A$_{f}$ solution is added. Thus the system approximately retains a constant volume so that $V\cong V_{B}$ throughout the entire branching process: $\chi_{1}=0$ and $\chi_{2}=1$.
 \end{enumerate}
 \end{minipage}}
%%%%%%%%%%%%%%%%%% Fig. (2)
\begin{wrapfigure}[12]{r}{5.5cm}
\vspace*{13mm}
\includegraphics[width=5.5cm]{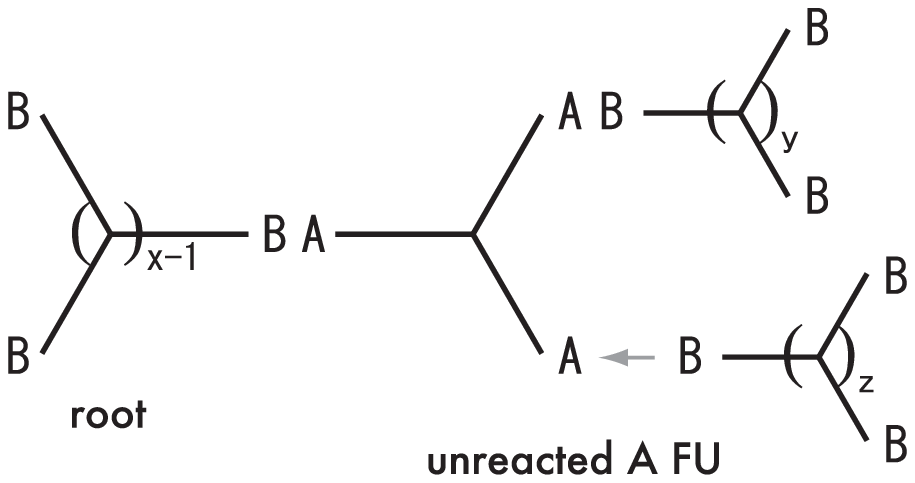}\\[3mm]
\setlength{\baselineskip}{10.5pt}{\small  Fig. 2: Representation of a typical \textit{n}-cluster formation from one R$-$A$_{f}$ monomer and three BB clusters.} 
\end{wrapfigure}
%%%%%%%%%%%%%%%%%%
\subsection{Gel Point Estimation}
The branching probability, $\alpha$, is defined as the probability that a A FU leads to the next A FU. Let $p_{\hspace{-0.2mm}R_{B}}$ be the fraction of cyclic bonds to all possible bonds for the R$-$B$_{g}$ monomer so that $p_{\hspace{-0.2mm}R_{B}}=N_{R}(D_{B})/gM_{B}$; let $p_{\hspace{-0.2mm}R_{A}}$ be the corresponding quantity for the R$-$A$_{f}$ monomer. Fig. 2 shows a typical (BB$-$AA)$_{n}$BB cluster formation from three BB clusters and one R$-$A$_{3}$ monomer ($n=x-1+1+y+z$). In this case we put unreacted B FU's on the root. Note that cyclization is possible only for unreacted A FU's on the R$-$A$_{3}$ monomer in question which are just added to the system for this unit reaction, contrary to the conventional homogeneous polymerization\cite{Kazumi}. $\alpha$, thus, should be written in the form:

 %%%%%%%%%%%%%%%%%% Eq. (39)
\begin{equation}
\alpha=D_{B}\sum_{m=0}^{f-1}m\binom{f-1}{m}\left(1-\frac{p_{\hspace{-0.2mm}R_{A}}}{D_{A}}\right)^{m}\left(\frac{p_{\hspace{-0.2mm}R_{A}}}{D_{A}}\right)^{f-1-m}\Big(g-1\Big).
\end{equation}
The gelation occurs at $\alpha=1$, which, with the equality $\frac{p_{\hspace{-0.2mm}R_{A}}}{D_{A}}\equiv \frac{p_{\hspace{-0.2mm}R_{B}}}{D_{B}}$, yields $$D_{B_{c}}=\frac{1}{(f-1)(g-1)}+\frac{N_{R}(D_{B_{c}})}{gM_{B}}.$$
For a large $n$, we can approximate that $N_{R}(D_{B_{c}})\fallingdotseq N_{R}(n_{c})$. And we have
%%%%%%%%%%%%%%%%%% Eq. (40)
\begin{equation}
D_{Bc}\fallingdotseq\frac{1}{(f-1)(g-1)}+\frac{N_{R}(n_{c})}{gM_{B}}.
\end{equation}
Eq. (40) is of the form:
%%%%%%%%%%%%%%%%%% Eq. (41)
\begin{equation}
D_{Bc}\equiv D(inter) + D(ring),
\end{equation}
as expected.

To apply eq. (35) beyond $D_{B}=D_{Bco}$, expand $N_{R}(n_{c})$ with respect to $D_{B_{c}}=D_{Bco}$ to yield
%%%%%%%%%%%%%%%%%% Eq. (42)
\begin{equation}
N_{R}(n_{c})\cong N_{R}(n_{co})+\frac{\Delta N_{R}(n_{co})}{\lambda}\left(D_{B_{c}}-D_{B_{co}}\right).
\end{equation}
We introduce the new quantity, the initial dilution ratio $r=\gamma_{f.i}/\gamma_{g,i}$. At $n=n_{co}$, since $n_{co}\,\lambda\,(f-1)(g-1)=1$, we have then
%%%%%%%%%%%%%%%%%% Eq. (43)
\begin{align}
\frac{\Delta N_{R}(n_{co})}{\lambda}&=\sum_{x=1}^{\infty}\varphi_{x}\frac{1}{2x}\Big(r+(f-1)(g-1)\Big)\,n_{co}\left\{1-\Big(\frac{n_{co}-1}{n_{co}}\Big)^{x}\right\}\,V_{B}\nonumber\\
&\equiv\frac{r+(f-1)(g-1)}{2}\sum_{x=1}^{\infty}\varphi_{x}\frac{A(x)}{x}\,V_{B};\\
%%%%%%%%%%%%%%%%%% Eq. (44)
N_{R}(n_{co})&=\frac{1}{(f-1)(g-1)}\sum_{x=1}^{\infty}\varphi_{x}\frac{1}{2x}\left\{r\left(1-\sum_{k=0}^{n_{co}-1}\frac{k^{x}}{n_{co}^{x+1}}\right)+(f-1)(g-1)\right\}\,V_{B}\nonumber\\
&\equiv\frac{1}{(f-1)(g-1)}\sum_{x=1}^{\infty}\varphi_{x}\frac{B(x)}{2x}\,V_{B},
\end{align}
where 
\begin{align}
A(x)&=\displaystyle n_{co}\left\{1-\Big(\frac{n_{co}-1}{n_{co}}\Big)^{x}\right\},\nonumber\\
B(x)&=\displaystyle\left\{r\left(1-\sum\nolimits_{k=0}^{n_{co}-1}\frac{k^{x}}{n_{co}^{x+1}}\right)+(f-1)(g-1)\right\}.\nonumber
\end{align}
Substituting eqs. (42)-(44) into eq. (40), we have
%%%%%%%%%%%%%%%%%% Eq. (45)
\begin{equation}
D_{Bc}=\frac{1}{(f-1)(g-1)}\left\{\frac{\displaystyle1-\left(\frac{r+(f-1)(g-1)}{2}\sum\nolimits_{x=1}^{\infty}\varphi_{x}\frac{A(x)}{x}-\frac{1}{2}\sum\nolimits_{x=1}^{\infty}\varphi_{x}\frac{B(x)}{x}\right)\cdot\gamma_{g,i}}{\displaystyle1-\frac{r+(f-1)(g-1)}{2}\sum\nolimits_{x=1}^{\infty}\varphi_{x}\frac{A(x)}{x}\cdot\gamma_{g,i}}\right\}.
\end{equation}
Eq. (45) is a general expression of the gel point in the drop-wise addition polymerization. Unfortunately the solution is not very easy to use, since it contains double sum of \textit{x} and \textit{k}. So, it is more convenient to approximate eq. (45) by the limiting case of $n=\infty$ (see eq. (46)). Fortunately this is possible, because $n$ is often very large and the difference between eq. (45) and eq. (46) is almost negligible.\\[3mm]

%%%%%%%%%%% Remark 3
\colorbox[gray]{0.90}{
\begin{minipage}{0.94\textwidth}
\paragraph{Remark 3:}
As $n\rightarrow\infty$, $A(x)\rightarrow x$ and $B(x)\rightarrow \left\{\displaystyle r\left(\frac{x}{x+1}\right)+(f-1)(g-1)\right\}.$ Then eq. (45) leads to
%%%%%%%%%%%%%%%%%% Eq. (46)
\begin{equation}
D_{Bc}=\frac{1}{(f-1)(g-1)}\left\{\frac{\displaystyle1-\left(\frac{r}{2}\sum\nolimits_{x=1}^{\infty}\varphi_{x}\Big(1-\frac{1}{x+1}\Big)+\frac{(f-1)(g-1)}{2}\sum\nolimits_{x=1}^{\infty}\varphi_{x}\Big(1-\frac{1}{x}\Big)\right)\cdot\gamma_{g,i}}{\displaystyle1-\frac{r+(f-1)(g-1)}{2}\sum\nolimits_{x=1}^{\infty}\varphi_{x}\cdot\gamma_{g,i}}\right\}.
\end{equation}
Now the gel point is a function of the dilution ratio, $r$, and the initial dilution, $\gamma_{g,i}$.
Since $r$ is given as an experimental condition, the gel point is calculable from the first principle.
 
\hspace{3mm}For $\gamma_{g,i}\gg\gamma_{f,i}\hspace{1.5mm} (\text{that is},\,\text{for}\,\, r\rightarrow 0)$, eq. (46) reduces to
%%%%%%%%%%%%%%%%%% Eq. (47)
\begin{equation}
\lim_{r\rightarrow 0}\text{eq. (46)}\,\,\rightarrow\,\, D_{Bc}=\frac{1}{(f-1)(g-1)}\left\{\frac{1-\frac{(f-1)(g-1)}{2}\displaystyle\sum\nolimits_{x}\varphi_{x}(1-1/x)\hspace{0.5mm}\gamma_{g}}{1-\frac{(f-1)(g-1)}{2}\displaystyle\sum\nolimits_{x}\varphi_{x}\hspace{0.5mm}\gamma_{g}}\right\}.
\end{equation}
And in the limit of the infinite concentration, $\gamma_{g}\rightarrow 0$, we recover the classical relation:
%%%%%%%%%%%%%%%%%% Eq. (48)
\begin{equation}
\lim_{\gamma_{g}\rightarrow 0}\text{eq. (46)}\,\,\rightarrow\,\, D_{Bco}=\frac{1}{(f-1)(g-1)}.\\[1mm]
\end{equation}
\end{minipage}}\\[3mm]

\section{Comparison with Experiment}

%%%%%%%%%%%%%%%%%% Fig. (3)
\begin{wrapfigure}[17]{r}{8cm}
\vspace*{-10mm}
\includegraphics[width=8cm]{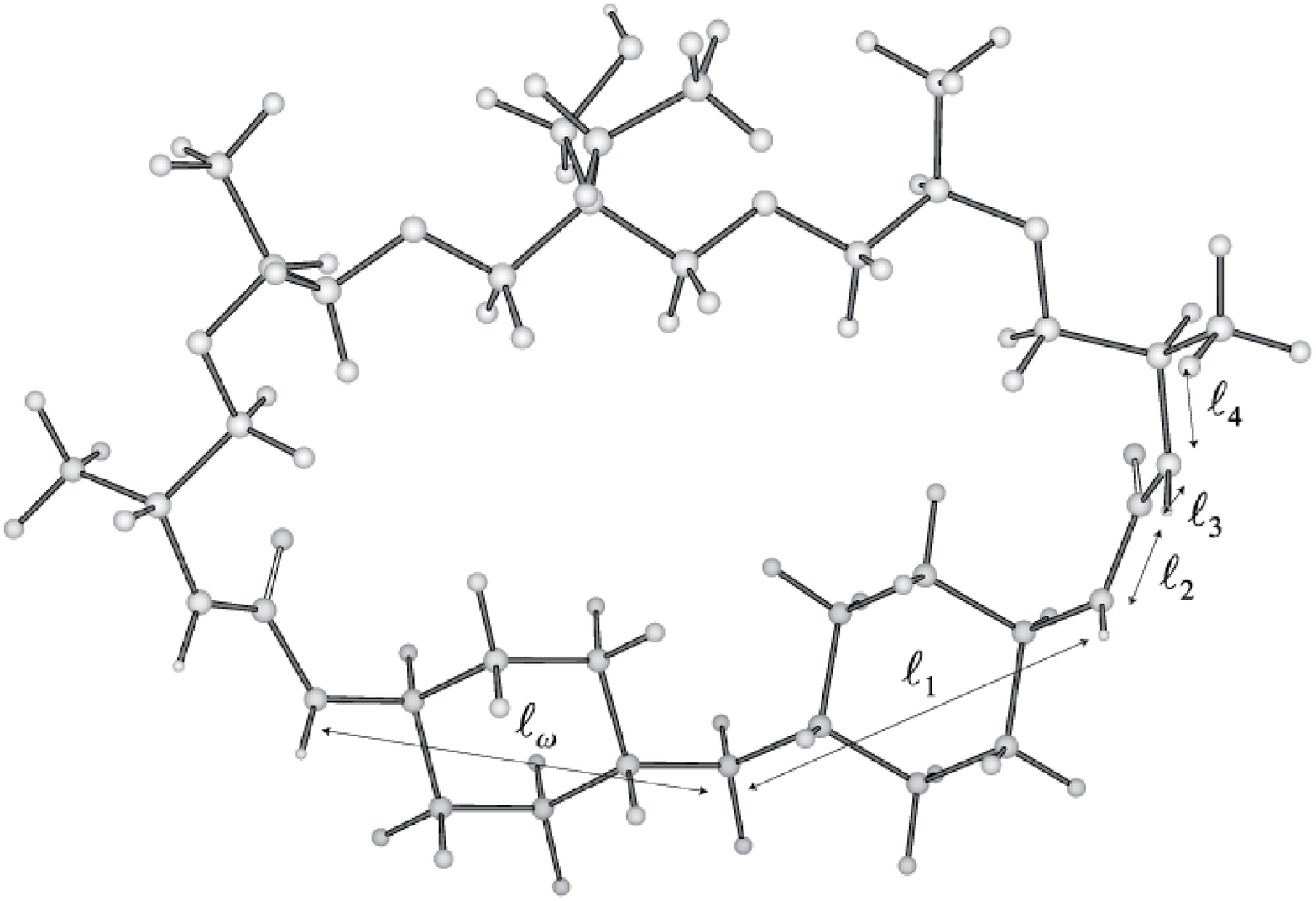} \\[2mm]
\setlength{\baselineskip}{10.5pt}{\small  Fig. 3: Representation of the smallest ring to be produced in the polymerization of HMDI and TRI. The dicyclohexane moiety is assumed to take the equatorial-equatorial conformation.} 
\end{wrapfigure}

To evaluate $D_{B_{c}}$ as a function of $\gamma_{g,i}$, we must calculate the cyclization frequency: $\varphi_{x}$. In concentrated solutions, we can expect the ideal behavior of branched molecules (no ring formation and no excluded volume), as mentioned earlier\cite{Kazumi}. Then $\varphi_{x}$ can be expressed by the incomplete gamma function of the form:
%%%%%%%%%%%%%%%%%% Eq. (49)
\begin{equation}
\varphi_x=\big(d/2\pi^{d/2}\ell_{s}^{\hspace{0.3mm}d}N_{A}\big)\displaystyle\int_{0}^{d/2\nu_{\hspace{-0.3mm}x}}\hspace{-2mm}t^{\frac{d}{2}-1}e^{-t}dt,
\end{equation}
where
%%%%%%%%%%%%%%%%%% Eq. (50)
\begin{equation}
\langle r_{x}^{2}\rangle_{\varTheta}=\nu_{x}\hspace{0.3mm}\ell_{\hspace{-0.3mm}s}^{\hspace{0.3mm}2}=C_{F}\hspace{0.2mm}\xi_{e}\hspace{0.3mm}x\hspace{0.3mm}\ell_{\hspace{-0.3mm}s}^{\hspace{0.3mm}2},
\end{equation}
\noindent is the end-to-end distance for an $x$-chain in the $\varTheta$ regime; $\ell_{s}$ denotes the length (1.37\hspace{0.3mm}\AA) of the N$-$C bond in the urea moiety ($\ell_{3}$ in Fig. 3), $C_{F}$ the Flory characteristic ratio, $\xi_{e}$ the effective bond number defined earlier\cite{Kazumi}, and $x$ the number of repeating units. 

In this paper, we take up the polyaddtion reaction of bis(4-isocyanatocyclohexyl)methane (HMDI: R$-$A$_{2}$ CH$_{2}$[(C$_{6}$H$_{10}$)NCO]$_{2} $) and poly(oxyalkylenetriamine) (TRI: R$-$B$_{3}$)\cite{Yilgor}. To seek the numerical estimate of $\varphi_{x}$, we must determine $C_{F}$ and $\xi_{e}$. 

There is little information about the expansion factor of poly(urea). It is important to notice that, according to eq. (46),  the location of the gel point depends only on the macroscopic quantity of cyclic species, namely, the total amount, but not on the microscopic detail of the distribution function. As a result, the quantity $C_{F}$ operates as an ajustment parameter for the total quantity of cyclic species. Then it is readily found through the numerical simulation that when we apply $C_{F}=4.5$, close to the value employed to the poly(urethane) homologue, a good result is obtained. The numerical estimate of 4.5 seems reasonable, but it is not yet conclusive; the validity should be verified by another experimental observations.

To calculate the effective bond number, $\xi_{e}$, let us examine the stereochemistry of the dicyclohexane moiety on HMDI. There are two known conformations of the cyclohexane ring convertible to each other, the skew and the chair. The chair form represents the lowest energy minimum, while the skew has higher energy because of the presence of the steric hindrance ($\Delta E=22.6\, k\hspace{0.3mm}J/mol$) due to the two axial $1,4$-hydrogen atoms\cite{Allinger}, with the statistical weight, exp\hspace{0.3mm}$\left(-\Delta E/RT\right)\approx10^{-4}$, showing that the cyclohexane ring exists, in equilibrium, exclusively in the chair form. Then consider the configuration of $1,4$-dimethyl  cyclohexane. According to the MM2 model calculation, there are three configurational states having energy minima that correspond to: equatorial-equatorial in which the two methyl moieties are splayed out ($\Delta E=0\,kJ$), equatorial-axial in which one methyl sticks up or down to the structure and the other is splayed out ($\Delta E=+7.29\,kJ$), and axial-axial in which both the methyls stick up and down to the structure ($\Delta E=+14.48\,kJ$). As expected, the equatorial-equatorial configuration is the most stable, and has the statistical weight of 18.68, which amounts to $\approx 95\,\%$ population of all configurations. Hence we may conclude that the dicyclohexane moiety exists almost exclusively in the equatorial-equatorial configuration.

From the above consideration, we obtain the imaginary bond length, $\ell_{1}\simeq 5.83 \text{\hspace{0.3mm}\AA}$.
Making use of this result, we can determine all the parameters (see Table 1). 

%%%%%%%%%%%%%%%%%% Table 1
\begin{table}[h]
\vspace{0mm}
\caption{Parameters for the HMDI-TRI Branched Poly(urea)}
\begin{center}
\begin{tabular}{c c c}\hline\\[-4mm]
parameters & \hspace{3mm}unit & \hspace{3mm}values \\[1mm]
\hline\\[-2mm]
Molecular Weight &  & \hspace{3mm} $\text{HMDI}=262$\\[1mm]
 &  & \hspace{3mm} $\text{TRI}=440$\\[1mm]
$f$ & \hspace{3mm} & \hspace{3mm} 2 \\[1mm]
$g$ & \hspace{3mm} & \hspace{3mm} 3 \\[1mm]
$d$ & \hspace{3mm} & \hspace{3mm} 3 \\[1mm]
$C_{F}$ & \hspace{3mm} & \hspace{3mm} 4.5\\[1mm]
$\xi_{e}$ & \hspace{3mm} & \hspace{3mm} 56\\[1mm]
$\ell_{s}$ & \hspace{3mm} $\left(\text{\AA}\right)$ & \hspace{3mm} 1.37 \\[2mm]
Cyclization Frequency & \hspace{3mm} $(mol/l)$ &  \\[1.5mm]
$\sum\nolimits_{x=1}^{\infty}\varphi_{x}$ & & \hspace{3mm} 0.135\\[2mm]
$\sum\nolimits_{x=1}^{\infty}\varphi_{x}\frac{1}{x}$ & & \hspace{3mm} 0.069\\[2mm]
$\sum\nolimits_{x=1}^{\infty}\varphi_{x}\frac{1}{(x+1)}$ & & \hspace{3mm} 0.039\\[2mm]
\hline\\[-7mm]
\end{tabular}
\end{center}
\end{table}
%%%%%%%%%%%%%%%%%% Table 2
\begin{table}[h]
\caption{Experimental Data in the Drop-wise Addition of HMDI onto TRI\cite{Yilgor}}
\begin{center}
\begin{tabular}{c c c c c c}\hline\\[-3.5mm]
HMDI & [NCO] & TRI & $[\text{NH}_{2}]$ & $r$ & Gel Point \\[0mm]
(mol/l) & (equiv/l) & (mol/l) & (equiv/l) & dilution ratio & ($D_{B_{c}}$)  \\[0.5mm]
\hline\\[-2mm]
0.80 & 1.60 & 0.47 & 1.41 & 0.88 & 0.591 \\[1mm]
0.63 & 1.26 & 0.37 & 1.12 & 0.89 & 0.625 \\[1mm]
0.47 & 0.93 & 0.28 & 0.83 & 0.89 & 0.648 \\[1mm]
0.31 & 0.61 & 0.18 & 0.55 & 0.90 & 0.717 \\[1mm]
0.23 & 0.46 & 0.14 & 0.41 & 0.89 & 0.803 \\[1mm]
0.15 & 0.30 & 0.09 & 0.27 & 0.90 & \textit{no gelation} \\[2mm]
\hline\\[-0mm]
\end{tabular}
\end{center}
\end{table}

With the help of the parameters of Table 1, together with the observed dilution ratio, $r=\gamma_{f,i}/\gamma_{g,i}=0.89$ (Table 2), we can plot eq. (46) as a function of $\gamma_{g,i}$. In Fig. 4, open circles represents experimental points by Unal and coworkers, and the solid line the theoretical line by eq. (46). The general trend of the theory is in good accord with the observations. 

%%%%%%%%%%%%%%%%%% Discussion
\section{Discussion}
\subsection{Interpretation of Results}

%%%%%%%%%%%%%%%%%% Fig. (4)
\begin{wrapfigure}[18]{c}{6cm}
\vspace*{-6mm}
\includegraphics[width=6cm]{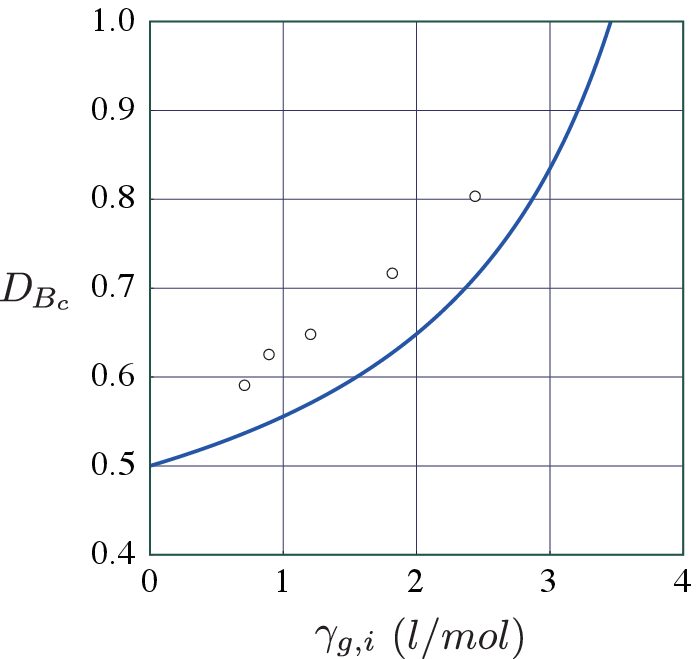} \\[2mm]
\setlength{\baselineskip}{10.5pt}{\small Fig. 4: Concentration dependence of gel points in the drop-wise addition reaction for $f=2$, $g=3$. \raisebox{-0.3mm}{\large$\circ$}\,: observed points by Unal and coworkers\cite{Yilgor}; solid line ($-$): theoretical line by eq. (46).} 
\end{wrapfigure}
As one can see, however there is appreciable numerical difference between the theory and the experiments. Fortunately this can be explained on the basis of (i) side reactions and (ii) deviation from the basic assumptions. 

It has been known earlier that isocyanates ($-\text{NCO}$) can react with alcohols ($-$OH) to form urethane bonds  
($-$NHCOO$-$). Hence, in the system cited above\cite{Yilgor}, the poly(urea) formation must always be accompanied by the urethane bond formation that wastes FU's because of the use of isopropyl alcohol as the reaction solvent. This necessarily shifts the gel point upwards. In light of the observations\cite{Stutz} of the cyclotrimerization of bisphenol-A dicyanate\,\footnote{\,There is a good example of the effect of side reactions on the shift of gel points\cite{Stutz}. The presence of H$_{2}$O moisture in the reactor is known to cause the hydrolysis of bisphenol-A dicyanate leading to the complicated side reactions and shifts the gel point upwards to a large extent, from the correct value 0.508 to 0.6 or higher values.}, it is probable that the alcoholysis of HMDI as a side reaction causes most of the discrepancy in question.

There might be other factors that cause the deviation. For instance, the theory has been derived on the assumption that the mixing is sufficient to assure the principle of equireactivity, while a set of reactions (1)-(3) must be completed within the minute interval of the unit cycle, so that the system always comprises (BB-AA)$_{n}$BB type molecules alone. This poses a problem because the system has to obey, on one hand, a very slow reaction with respect to the realization of the sufficient mixing; it has to obey, on the other hand, a very fast reaction with respect to the instantaneous completion of the reaction cycle (1)-(3); it is clear that a delicate balance is required to realize a genuine drop-wise polymerization, deviation from which should shift the gel point upwards.

Taking these circumstances into consideration, it is by no means unreasonable to conclude that there is a satisfactory agreement between the theory and the experiments.
\subsection{Comparison with Conventional Branching Process}
It will be of interest to inquire the question, 'If all R$-$A$_{f}$ molecules are added at once, where is the gel point observed\,?' The gel point in that conventional polymerization has been found to obey the equation\cite{Kazumi}:

%%%%%%%%%%%%%%%%%% Eq. (51)
\begin{equation}
D_{A_c}=\sqrt{\frac{1}{s}}\left\{\frac{1-\left(1+\kappa\right)\sqrt{s}\hspace{1mm}\displaystyle\sum\nolimits_{x}\left(1-1/2x\right)\varphi_{x}\hspace{0.3mm}\gamma}{1-\left(1+\kappa\right)\sqrt{s}\hspace{1mm}\displaystyle\sum\nolimits_{x}\varphi_{x}\hspace{0.3mm}\gamma}\right\},
\end{equation}
%%%%%%%%%%%%%%%%%%
where $s=(f-1)(g-1)/\kappa$, $\kappa=gM_{B}/nfL \,\,(\ge 1)$ as defined earlier, and $\gamma^{-1}=\frac{nfL+gM_{B}}{V}$. $\kappa$ corresponds to the reciprocal of the gel point in the drop-wise addition polymerization. By eq. (25) we have $V=V_{B}(1+r/\kappa)$. Substituting this into eq. (51) together with some rearrangement, we have
%%%%%%%%%%%%%%%%%% Eq. (52)
\begin{equation}
D_{B_c}=\sqrt{\frac{1}{s'}}\left\{\frac{1-\left(1+r/\kappa\right)\sqrt{s'}\hspace{1mm}\displaystyle\sum\nolimits_{x}\left(1-1/2x\right)\varphi_{x}\hspace{0.3mm}\gamma_{g,i}}{1-\left(1+r/\kappa\right)\sqrt{s'}\hspace{1mm}\displaystyle\sum\nolimits_{x}\varphi_{x}\hspace{0.3mm}\gamma_{g,i}}\right\},
\end{equation}
%%%%%%%%%%%%%%%%%%
where $s'=\kappa\,(f-1)(g-1)$, and $r=\gamma_{f,i}/\gamma_{g,i}$ and $\gamma_{g,i}$ the reciprocal of the initial B FU's concentration before mixing as defined in the text. Physically, $D_{B_{c}}$ in eq. (52) must be less than $1/\kappa=nfL/gM_{B}$ (the gel point in the drop-wise addition polymerization), because exactly $nL$ R$-$A$_{f}$ molecules are mixed at once with $M_{B}$ R$-$B$_{g}$ molecules. The calculation of eq. (52) showed that $D_{B_{c}}=0.595$ for $1/\kappa=0.591$, 0.632 for $1/\kappa=0.625$ and so forth. It was found that for all examples, the gel points ($D_{B_{c}}$) of the conventional polymerization exceed those ($1/\kappa$) of the drop-wise addition polymerization. This indicates that the gelation will never occur in the conventional polymerization if the reaction is carried out under the same conditions as those employed in the drop-wise addition polymerization. In order for the gelation to occur, more concentrated circumstances are needed. In other words, in the drop-wise addition polymerization, the system behaves as if the cyclization is less frequent than in the conventional branching process.
%%%%%%%%%%%%%%%%%% References

\end{document}